\documentclass{article}

\usepackage[authoryear]{natbib}

\usepackage[preprint]{neurips_2024}

\usepackage[utf8]{inputenc} 
\usepackage[T1]{fontenc}    
\usepackage{hyperref}       
\usepackage{url}            
\usepackage{booktabs}       
\usepackage{amsfonts}       
\usepackage{nicefrac}       
\usepackage{microtype}      
\usepackage{xcolor}         

\usepackage{amsmath}
\usepackage{float}

\usepackage{graphicx}
\usepackage{multirow}
\usepackage{tabularx}
\usepackage{array}
\usepackage{multicol}
\usepackage{algorithm}
\usepackage[noend]{algpseudocode}
\usepackage{siunitx}

\title{E3STO: Orbital Inspired SE(3)-Equivariant Molecular Representation for Electron Density Prediction}

\author{%
  Ilan Mitnikov \\
  Center of Bits and Atoms\\
  Massachusetts Institute of Technology\\
  Cambridge, MA 02139 \\
  \texttt{ilanm@mit.edu} \\
  \And
  Joseph Jacobson\\
  Center of Bits and Atoms\\
  Massachusetts Institute of Technology\\
  Cambridge, MA 02139 \\
}

\begin{document}

\maketitle

\begin{abstract}
  Electron density prediction stands as a cornerstone challenge in molecular systems, pivotal for various applications such as understanding molecular interactions and conducting precise quantum mechanical calculations. However, the scaling of density functional theory (DFT) calculations is prohibitively expensive. Machine learning methods provide an alternative, offering efficiency and accuracy. We introduce a novel SE(3)-equivariant architecture, drawing inspiration from Slater-Type Orbitals (STO), to learn representations of molecular electronic structures. Our approach offers an alternative functional form for learned orbital-like molecular representation. We showcase the effectiveness of our method by achieving SOTA prediction accuracy of molecular electron density with 30-70\% improvement over other work on Molecular Dynamics data.
\end{abstract}

\section{Introduction}

The problem of electron density prediction involves determining the distribution of electron probability density of a molecule. This distribution is crucial for understanding chemical properties and reactivity. Accurate electron density prediction requires solving the Schrödinger equation for many-electron systems, which is computationally challenging due to the complex interactions between electrons. Traditional methods like Hartree-Fock and Density Functional Theory (DFT) are commonly used, but they involve approximations that can affect the accuracy of the predicted electron densities. Recent advances in machine learning and quantum computing offer promising approaches to improve the precision and efficiency of these predictions \cite{jorgensen2020deepdft},\cite{bogojeski2018efficient}.

Previous works have explored various machine learning models for electronic structure prediction. SchNet \cite{schutt2018schnet} is a notable early example that laid the groundwork for applying neural networks to atomistic systems. These models, however, often lack rotational equivariance, which is essential for capturing the symmetries inherent in molecular systems. Equivariant neural networks, such as Tensor Field Networks (TFN) \cite{thomas2018tensor} and SE(3)-equivariant networks, have been developed to address this limitation by incorporating rotational and translational symmetries directly into the model architecture. These advancements ensure that the predictions are handled properly under rotations and translations of the input structures, leading to more accurate and generalizable models.

In this paper, we introduce electron density prediction method that integrates the mathematical form of Slater Type Orbitals (STOs) with SE(3)-equivariant neural networks. STOs have been extensively used in quantum chemistry due to their realistic representation of atomic electron distributions. In particular, our method involves dynamically generating STO-inspired basis functions for each individual molecule. We take a new approach where the functional form of the basis functions themselves depends on molecule we want to represent.  By leveraging the mathematical properties of STOs and the symmetry-preserving capabilities of SE(3)-equivariant networks, we achieve state-of-the-art performance in electron density prediction, opening new avenues for accurate and efficient quantum mechanical modeling.

\section{Related Work}
\label{rel_work}

\subsection*{2.1 Machine Learning for Electron Density Estimation}

Recent advancements in machine learning (ML) present promising solutions to mitigate the computational challenges inherent to electron prediction tasks. By leveraging data-driven models, predicting electronic properties can be achieved at a significantly reduced computational cost. Neural networks, in particular, have demonstrated great potential in approximating complex quantum mechanical properties. For instance, \cite{sinitskiy2018deep} employed a voxel-based 3D U-Net convolutional network architecture to predict electron densities at the voxel level. Additionally, \cite{jorgensen2020deepdft}, \cite{cheng2023equivariant}, and \cite{kim2024gaussian} have approached electron density prediction using neural message passing on graphs, embracing multicentric approximation framework and developing an equivariant graph neural network (GNN) based on tensor products for predicting density spectra. Other notable works include \cite{unke2021se}, where they directly modeled the orbitals as part of a Hamiltonian prediction task.
And, \cite{brockherde2017bypassing} combined DFT calculations with machine learning to accelerate electron density predictions. A concurrent work, \cite{fu2024recipechargedensityprediction} provided a scheme which achieves SOTA accuracy on the QM9 dataset \cite{ruddigkeit2012enumeration} by adding virtual orbitals which are not centered at atomic positions. They follow a similar approach to ours and use SE(3)-equivariant neural networks.

\subsection*{2.2 SE(3)-Equivariant Networks}

Equivariant neural networks have been developed to address the limitations of rotational invariance in molecular modeling. \cite{thomas2018tensor} introduced Tensor Field Networks (TFNs), which use SO(3) equivariant convolutions to maintain rotational symmetry. These networks have shown improved performance in various tasks involving 3D data.

Building on this, \cite{unke2021se} proposed SE(3)-equivariant neural networks specifically designed for molecular wavefunctions and electron densities, demonstrating significant improvements in prediction accuracy. These networks leverage the inherent symmetries of physical systems, making them well-suited for modeling molecular structures and interactions.

\section{Methods}
\subsection{Atomic and Molecular Orbitals}
\label{orbitals}
Atomic orbitals serve as the foundational building blocks for constructing molecular orbitals, which describe the quantum states of electrons in a molecule. These molecular orbitals are calculated using the Linear Combination of Atomic Orbitals (LCAO), where each molecular orbital is expressed as a weighted sum of atomic orbitals \cite{pauling1931nature}. The coefficients in this linear combination are determined by solving the Roothaan-Hall equations, which involve computing the Fock matrix, overlap integrals, and iteratively refining the orbital coefficients \cite{roothaan1951new}. 

Slater Type Orbitals (STOs) are useful for this purpose due to their exponential radial decay and accurate representation of electron distributions. Each electron's wavefunction can be decomposed into a radial part and an angular part, the latter being described by spherical harmonics. An STO of an electron in an atom is given by the equation \cite{slater}:

\begin{equation}
\chi_{nlm}(r, \theta, \phi) = N r^{n-1} e^{-\zeta r} Y_\ell^m(\theta, \phi)
\end{equation}

where \( N \) is a normalization constant, \( n \) is the principal quantum number, \( \zeta \) is the orbital exponent related to the effective nuclear charge, and \( Y_{\ell m}(\theta, \phi) \) are the spherical harmonics that are indexed by the angular quantum numbers \( \ell \) and \( m \). These orbitals are crucial in quantum chemistry for providing accurate descriptions of atomic and molecular systems. The total molecular orbital is then computed as a linear combination of atomic orbitals (LCAO) \cite{pauling1931nature}: 
\begin{equation}
    \psi_i(\mathbf{r}) = \sum_k c_{ki} \chi_k(\mathbf{r})
\end{equation}
where $\psi_i$ are the molecular orbitals,
\begin{equation}
    \rho(\mathbf{r})=\sum_{i,j} q_{ij} \psi_i(\mathbf{r})\psi_j(\mathbf{r})=\sum_{i,j,k,p} q_{ij} c_{ki} c_{pj} \chi_k(\mathbf{r}) \chi_p(\mathbf{r})
\end{equation}
 and $\rho$ is the total molecular electron density.
\subsection{Molecule Dependent Expansion}
\label{sec:basis}
The natural representation of atomic orbitals as spherical harmonics, which carry different symmetries provides a good basis for parameterizing these physical systems with equivariant neural networks that follow similar symmetries. We introduce a continuous representation that is  learnable and specific to each molecule. Unlike traditional basis sets that are fixed and predefined, our approach generates basis functions dynamically. The basis functions are optimized during the training process to best represent the electronic structure of the given molecule. As a first step, we follow the LCAO atom-centred expansion of molecular orbitals:

\begin{equation}
    \Psi(\mathbf{r};\mathbf{p}_\text{mol}) = \sum_{i \in \textit{Atoms}} \psi_i(\mathbf{r}-\mathbf{r_i}; \mathbf{p}_\text{mol})
\end{equation}

where $\psi_i$ are functions centered at the atoms and are dependent on the atoms of the molecule  $\mathbf{p}_\text{mol}=\{(a_i,r_i)\}_{i\in Atoms}$, where $a_i, r_i$ are the atomic elements and positions, respectively. In particular, inspired by STOs we define:
\begin{align}
    \psi_i(\mathbf{r}-\mathbf{r_i}; \mathbf{p}_\text{mol}) = & \sum_n \sum_\ell \mathcal{R}^i_{n\ell}(|\mathbf{r}-\mathbf{r_i}|; \mathbf{p}_\text{mol}) \times \mathcal{C}^i_{n\ell}(\mathbf{p}_\text{mol}) \cdot Y_{\ell}\left(\frac{\mathbf{r}-\mathbf{r_i}}{|\mathbf{r}-\mathbf{r_i}|}\right) 
    +\mathbf{MLP}_n^i(|\mathbf{r}-\mathbf{r_i}|; a_i)
\end{align}
where the radial function is inspired by STOs as follows:
\begin{equation}
\label{eq:rad}
     \mathcal{R}^i_{n\ell}(\mathbf{r}-\mathbf{r_i}; \mathbf{p}_\text{mol}) = f_{n\ell}(|\mathbf{r}-\mathbf{r_i}|; \mathbf{p}_\text{mol}) 
     \times 
     |\mathbf{r}-\mathbf{r_i}|^{\eta_{n\ell}(\mathbf{p}_\text{mol})}
    \times 
    e^{-\zeta_{n\ell}(\mathbf{p}_\text{mol}) |\mathbf{r}-\mathbf{r_i}|}
\end{equation}
and $f_{n\ell}, \eta_{n\ell}, \zeta_{n\ell}, \mathcal{C}_{n\ell}$ are functions parameterized by equivariant neural networks. The output of $\mathcal{C}_{n\ell}$ is represented by the same $SO(3)$ irreducible representation as $Y_\ell$. The $\mathbf{MLP}_n$ term accounts for potential features per atom type, which are not necessarily dependent on molecular structure. This mimics the common initial guess used in DFT calculation of non-interacting atomic orbitals.
The total electron density is calculated as:
\begin{equation}
    \rho(\mathbf{r}; \mathbf{p}_\text{mol}) = ||\Psi(\mathbf{r};\mathbf{p}_\text{mol})||^2
\end{equation}

In particular, we have introduced a dependence of the expansion parameters on the molecular structure $\mathbf{p}_\text{mol}$.
The functional form of $\psi_i$ was constructed to be analogous to STO functions:
\begin{equation}
\begin{aligned}
    \mathcal{C}^i_{n\ell} \quad &\leftrightarrow \quad c_{ki} \quad &\text{(Atomic Orbital Coefficients)} \\
    \eta_{n\ell} \quad &\leftrightarrow \quad n \quad &\text{(Principle Quantum Number)} \\
    \zeta_{n\ell} \quad &\leftrightarrow \quad \zeta \quad &\text{(Orbital Exponent)} \\
    f^i_{n\ell}f^j_{n\ell} \quad &\leftrightarrow \quad q_{ij} \quad &\text{(Density Matrix Coefficients)} \\
\end{aligned}
\label{eq:params}
\end{equation}
\textbf{Interpretation and Intuition}  Following the formulation and reasoning for molecular orbital expansion, we emphasize our distinction from other approaches. In our method, the coefficients in Eq. \ref{eq:params} depend on the molecular structure, unlike traditional approaches where these coefficients are specifically crafted for different quantum mechanical calculations. We argue that the traditional method limits the ability to directly approximate similar densities for similar molecules. Instead, we use atom-centered orbitals and spherical harmonic expansion (which mimics atomic orbitals) while allowing the radial component to be entirely learned and input-dependent. This approach strikes a balance between physics-based inductive bias and learnability.

Furthermore, our architecture is driven by the observation that only the atomic orbital coefficients $\mathcal{C}$ possess higher-order symmetries and are thus represented by higher-order spherical harmonics. This is intuitive as it reflects the natural behavior of atomic orbitals. In contrast, the radial coefficients $\eta$, $\zeta$, and $f$ are scalar quantities and invariant under any transformations. To ensure a meaningful representation for these parameters, we allow higher-order features to inform these coefficients. Our architecture endows these scalars with geometric information, as described below.

\subsection{Architecture}
We utilize 3D point cloud convolutions and tensor products as defined in Tensor Field Networks (TFN) \cite{thomas2018tensor}. In particular, the input is a molecule $\mathbf{p}_\text{mol}$ represented as a set of atomic elements $a_i \in \mathbb{Z}$ and 3D positions $r_i \in \mathbb{R}^3$ where $\mathbf{p}_\text{mol}=\{a_i,r_i\}_{i\in atoms}$. The position of each atom is defined relative to the mean position of all atoms. Furthermore, we do not rely on any other knowledge of molecular structure, such as bonds. To incorporate the fact that molecular electron density is a $SO3$ equivariant property we represent each molecule with a set of equivariant features. The initial parametrization of a molecule using atomic elements and positions naturally corresponds to a set of $SO3$ equivariant irreducible representation of order $\ell=0$ and $\ell=1$, respectively. To be explicit, let the initial $SO3$ irreducible representation of the molecule is $\mathbf{V}_{init} =\{V_i^{\ell=0},V_i^{\ell=1}\}_{i\in atoms}=\{a_i,r_i\}_{i\in atoms}=\mathbf{p}_\text{mol}$. However, atomic orbital theory suggests that to represent molecular electron density of arbitrary complexity, higher frequency components, represented by higher order ($\ell>1$) spherical harmonics, are needed. 
\begin{figure}
    \centering
    \includegraphics[width=1.0\linewidth]{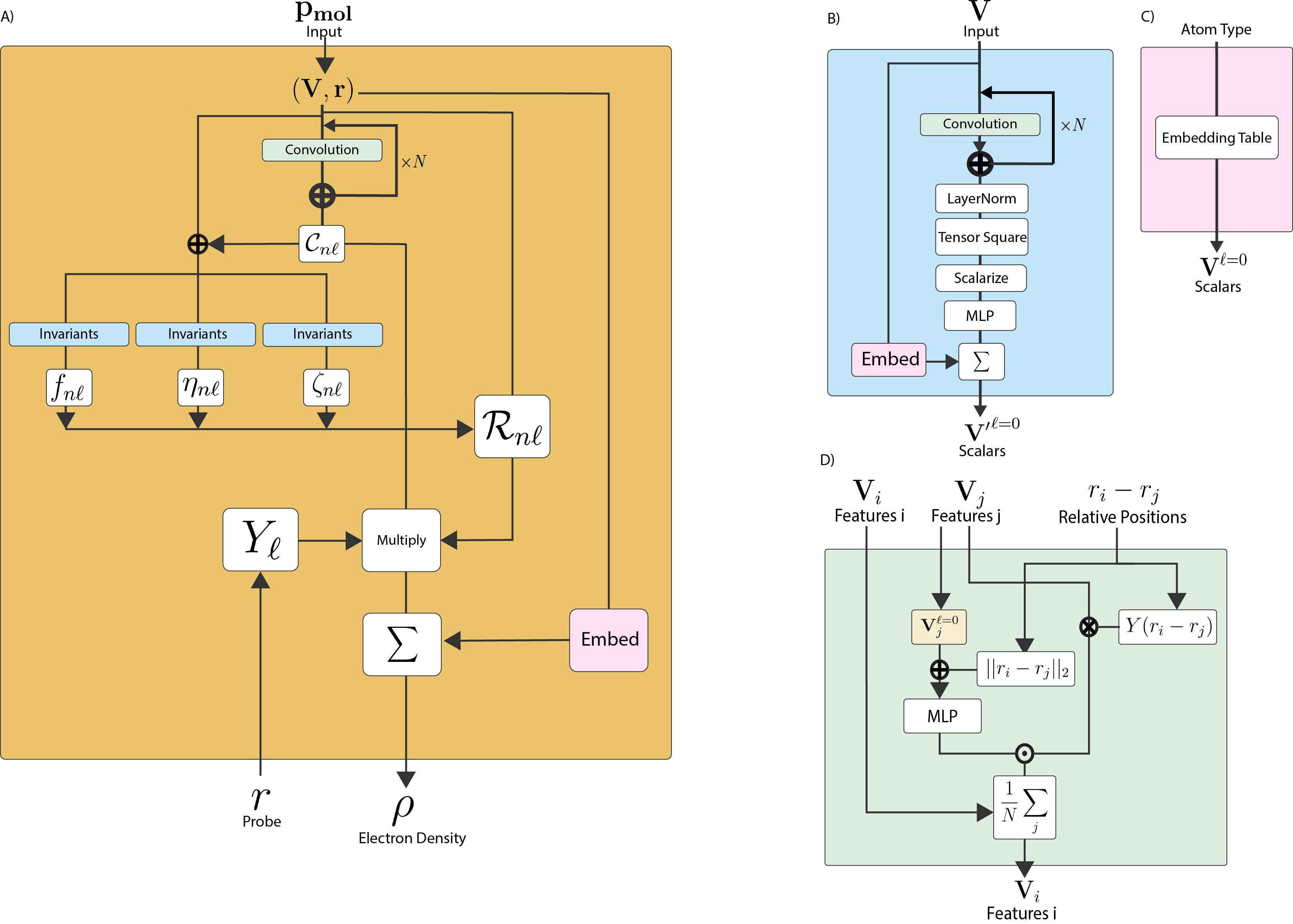}
    \caption{A) Schematic of total electron density prediction given molecular structure. B) $Invariants$ algorithm for extracting meaningful invariant features from higher order equivariant features. C) $Embed$ function which uses a simple lookup table by atom type. D) $Convolution$ which mixes features of neighboring atoms and creates richer representation for the molecular orbitals.}
    \label{fig:e3sto}
\end{figure}

\textbf{Convolution} converts the initial $\ell\leq 1$ into higher order features using the equivariant 3D point cloud convolution as introduced in TFN. In principle, higher order features are obtained by applying \textit{generalized tensor product} operations between features of neighboring atoms and the . Let $u \in atoms $,  the set of neighboring atoms $v \in \mathcal{N}(u)$ is defined by the K-nearest neighbor (KNN) algorithm (with $K=8$). In particular, after one round of convolution the updated higher order representation of atom $u$ is given by:

\begin{equation}
\label{eq:tp}
    \mathbf{V'}_{u}^{\ell'} = \sum_{v \in \mathcal{N}(u)} \sum_{\ell \geq 0} \mathbf{W}^{\ell'}_{\ell}(r_v - r_u) \mathbf{V}_{v}^{\ell} + \mathbf{V}_{u}^{\ell'}
\end{equation}
where $\mathbf{W}^{\ell'}_{\ell}(r_v - r_u)$ are the coefficients determined by the \textit{Clebsch-Gordon tensor product} and by learned radial components:  
\begin{equation}
\mathbf{W}^{\ell'}_{\ell}(\mathbf{r}) = \sum_{k=0}^J \varphi^{\ell'}_{\ell k}(r) \sum_{m=-k}^{k} Y_{k}^{m}(\hat{\mathbf{r}}) C^{\ell'}_{\ell k}
\end{equation}
where $C_{\ell k}^{\ell'}$ is the Clebsch-Gordan coefficients which are non-zero for $\ell' \in [|\ell-J|,\ell+J]$. Furthermore, $\varphi_{\ell k}^{\ell}: \mathbb{R}^{+} \rightarrow \mathbb{R}$ are learnable radial nets, and $J$ is the spherical harmonic embedding order of the relative position between neighboring atoms (in this work $J=2$). The convolution operation defined above is $SO3$ equivariant as shown in \cite{thomas2018tensor}.  One can immediately notice that, when starting with $\ell_{max}=1$ as features, by applying N rounds of convolution the highest representation order becomes $\ell_{max}=1+JN$. Finally, we apply an equivariant linear layer which mixes features of the same order and $n$ different channels:
\begin{equation}
    \mathbf{V_n'}_{u}^{\ell} = \sum_m A_{nm}^\ell \mathbf{V_m}_{u}^{\ell}
\end{equation}
Please refer to Alg. \ref{alg2} for pseudo-code, we use $\oplus$ to refer to concatenation and $\otimes$ to tensor products. (Fig. \ref{fig:e3sto} D)The equivariant layers are implemented using \textit{e3nn-jax} \cite{geiger2022e3nn}. Where the point cloud convolution use spherical harmonic embeddings (of order $J$) of displacement vectors to enhance the representation. Fully connected tensor products with learnable weights are used as part on the convolutional layers, as described in \cite{geiger2022e3nn}. 

\begin{minipage}{0.50\textwidth}
\begin{algorithm}[H]
    \centering
    \caption{Convolution}\label{alg2}
    \footnotesize
    \begin{algorithmic}[1]
    \Require $\{r_i\}_i, \mathbf{V}$
    \For{$i \in 1 \dots \mathbf{N}$} 
        \State $ \vartheta_{ij} \gets \mathbf{V}_j \otimes Y_k(r_i-r_j) $
        \State $\varphi_{ij} \gets \varphi(|r_i-r_j|)$
        \State $\mathbf V_i \gets \mathbf V_i + \sum_{j \in \mathcal N_i}\vartheta_{ij} \circ MLP(\varphi_{ij} \oplus V_i^{\ell=0})$  
        \EndFor
    \State $\mathbf{V} \gets EquivariantLinear(\mathbf{V})$
    \State \Return $\mathbf{V}$
    \end{algorithmic}
\end{algorithm}
\end{minipage}

\textbf{Orbital Coefficients} 
As described in Sec. \ref{orbitals}, atomic orbitals are represented by a combination of spherical harmonics $Y_\ell$ which are $SO3$ equivariant functions. To predict molecular electron density, we aim to learn a suitable combination of spherical harmonics for each atom, which are represented by a set of equivariant features $\mathbf{V} = \{\mathbf{V}^\ell\}_{\ell \in [0,\ell_{max}]}$. Given an initial molecular parameterization $\mathbf{p}_{mol}=\mathbf{V}=\{V_i^{\ell=0},V_i^{\ell=1}\}_{i\in atoms}$ we apply $N$ convolution rounds (Alg. \ref{alg2}) to obtain $\mathbf{V'_n} = \{\mathbf{V'_n}^\ell_i\ | i\in atoms, \ell \in [0,1+JN]\}$. We denote the resulting features $\mathcal{C}_{nl}=\mathbf{V'_n}^\ell$, which can be viewed as the learned expansion of atom-centered orbitals with $n$ basis functions (spherical harmonics) of order $\ell \in [0, \ell_{max}]$. In particular, the molecular orbital of order $\ell$ centred at atom $i$ is given by $\psi_{\ell}^i(r)=\sum_n \mathcal{R}^i_{n \ell}(r)\sum_{m=-\ell}^\ell \mathcal{C}^i_{n\ell m} Y^m_\ell (r)$. Where $\mathcal{R}^i_{n \ell}$ is the radial component described below.

\textbf{Radial Component} $R_{n\ell}$ as described in Eq. \ref{eq:rad} is a function of three components: $f_{n\ell},\eta_{n\ell},\zeta_{n\ell} \in \mathbb{R}^{|atoms|}$. Each component takes as input $\mathbf{V}=\mathbf{p_{mol}}$ and $\mathcal{C}_{n\ell}$ and outputs one scalar per atom and has the following form:
\begin{equation} Invariants(\mathbf{V},\mathcal{C}_{n\ell}) + Embed(\mathbf{V})
\end{equation}
Where $Invariants$ are given by Alg. \ref{alg3} (Fig. \ref{fig:e3sto}B):

\begin{minipage}{0.50\textwidth}
\begin{algorithm}[H]
    \centering
    \caption{Invariants}\label{alg3}
    \footnotesize
    \begin{algorithmic}[1]
        \Require $\mathbf V, \mathcal C_{nl}$
        \State $\mathbf V \gets \mathbf V \oplus \mathcal C_{nl}$ 
        \State $\mathbf V \gets Convolution^N(\mathbf{V})$ 
        \State $\mathbf V \gets Normalize(\mathbf{V})$ 
        \State $\mathbf S \gets Scalarize(\mathbf{V} \otimes \mathbf{V})$ 
        \State $\mathbf S \gets MLP(\mathbf{S})$ 
        \State \Return $\mathbf{S}$
    \end{algorithmic}
\end{algorithm}
\end{minipage}

The idea here is to derive invariant features from higher order features using the tensor products from the point cloud convolutions. This means that the geometric information of the atomic positions affects the learned basis coefficients. This is possible since each convolution round generates orders $\ell' \in [|\ell-J|,\ell+J]$, as described above. Thus, information flows both to higher and low order features. After a few round of convolution, to obtain scalar invariants from higher order tensor we project the representation to scalar by keeping only $l=0$ features,  $Scalarize(\mathbf{V})=\mathbf{V}^0$.

\textbf{Embedding} The $Embed$ function is a lookup table based on atom type (Fig. \ref{fig:e3sto}C). We use three embedding tables: one for each of $\eta_{n\ell}$, $\zeta_{n\ell}$, and $f_{n\ell}$. Additionally, we use an embedding table $Embed_{\rho}$ for the total density $\rho$ as a bias term for each atomic type (as used in line 8 of Algorithm \ref{alg1}). Each $Embed$ function takes the atom type $a_i$ as input and outputs a scalar. We apply this for each invariant/scalar coefficient that describes the orbital expansion as defined in Eq. \ref{eq:rad}. These embeddings account for any systematic deviations not captured by other parameters and constant features of specific atoms. This approach is commonly used in DFT calculations, where non-interacting atomic orbitals are used as the initial approximation.


\textbf{Total Electron Density} $\rho(\mathbf{r})$ is calculated by Algorithm \ref{alg1} (Fig. \ref{fig:e3sto}A). To compute the electron density, we first obtain a higher-order representation of atom-centered basis functions by applying $N$ rounds of convolution. This rich representation is then fed into three different $Embed$ and $Invariants$ networks, which are equivariant networks that further compute the coefficients of the molecular orbital expansion. After all the coefficients are computed by the neural network, they are combined according to the function described in Eq. \ref{eq:rad}. For calculating the total orbital centered at atom $i$, as described in line 7, we take the dot product between the higher-order tensor of the atomic representation $\mathcal{C}{n\ell}$ and the representation of the same order of the displacement vector relative to that atom, as given by the spherical harmonic embedding $Y\ell(r-r_i)$. This process allows us to obtain the electron contribution of each atom at each point in space. 

\begin{minipage}{0.54\textwidth}
\begin{algorithm}[H]
    \centering
    \caption{Electron Density $\rho(\mathbf{r})$}\label{alg1}
    \footnotesize
    \begin{algorithmic}[1]
        \Require $\mathbf{p_{mol}}=(a_i,r_i)$
        \State $\mathcal{C}_{n\ell} \gets Convolution^N(\mathbf{p_{mol}})$ 
        \State $f_{n\ell} \gets Invariants(\mathbf{p_{mol}}, \mathcal{C}_{n\ell}) + Embed_f(a_i)$ 
        \State $\mathcal{\eta}_{n\ell} \gets ||Invariants(\mathbf{p_{mol}}, \mathcal{C}_{n\ell}) + Embed_{\eta}(a_i)||$ 
        \State $\mathcal{\zeta}_{n\ell} \gets ||Invariants(\mathbf{p_{mol}}, \mathcal{C}_{n\ell}) + Embed_{\zeta}(a_i)||$ 
        \State $\mathbf{r}^i \gets \mathbf{r}-r_i$
        \State $\mathcal{R}^i_{n\ell} \gets f_{n\ell}(\mathbf r^i)\times ||\mathbf r^i||^{\mathcal{\eta}_{n\ell}} \times e^{-\mathcal{\zeta}_{n\ell}||\mathbf{r}^i||}$
        \State $\psi_i \gets \sum_{n,\ell} \mathcal{R}^i_{n\ell} ( \mathcal{C}_{n\ell} \cdot Y_\ell(\hat{\mathbf{r}}^i))$
        \State $\psi_i \gets \psi_i + MLP(||\mathbf{r}^i|| \oplus Embed_{\rho}(a_i))$
        \State \Return $||\sum_i \psi_i||^2$
    \end{algorithmic}
\end{algorithm}
\end{minipage}

 One can refer to Fig. \ref{fig:e3sto} for a schematic representation of the Alg. \ref{alg2},\ref{alg3},\ref{alg1}. We use tensor-product layers as implemented in \textit{e3nn-jax} \cite{geiger2022e3nn}, and $0e+1o+2e+3o+4e+5o+6e$ are the $irreps$ for our trials  with dimensions ranging from 4 to 8. 
 
 \textbf{Complexity Analysis} The convolution operation is base on K-nearest neighbors (KNN) algorithms. Let $N$ be the number of atoms in a molecule. The worst case is $O(NK)$ in terms of number of atoms. The tensor operation is quadratically dependent on of the dimension of representation and quadrupally on the degree $l_{max}$ $O(d^2l^4_{max})$. So overall supposing the most expensive operations of tensoring all neighbors everytime we get $O(NKd^2l^4_{max})$. In pratice $K,d,l$ stay fixed and only $N$ grows with the system.
 


\section{Experiments}
\subsection{Datasets and Baselines}
We evaluate our model on commonly used datasets for the task of electron density prediction. In particular, we follow closely two recent works for evaluation and as direct comparison for our method. The datasets and baseline information was gathered by \cite{cheng2023equivariant} and \cite{kim2024gaussian}. We use the following three datasets, each contains molecular geometries and electron density volumetric data on a 3D grid. 

\textbf{QM9.} The QM9 dataset \cite{ruddigkeit2012enumeration} includes 133,885 species with up to nine heavy atoms (CONF). The density data and the data split come from \cite{cheng2023equivariant}, yielding 123,835 training samples, 50 validation samples, and 10,000 testing samples.


\textbf{MD.} The dataset consists of 6 small molecules (ethanol, benzene, phenol, resorcinol, ethane, malonaldehyde) with different geometries sampled from molecular dynamics (MD). The first 4 molecules are from \cite{bogojeski2018efficient} with 1,000 sampled geometries each. The latter 2 are from \cite{bogojeski2020quantum} with 2,000 sampled geometries each. The models were trained separately for each molecule.

\textbf{Cubic}  This large-scale dataset contains electron densities on 17,418 cubic inorganic materials provided by \cite{wang2022large}. We follow the filtering of \cite{cheng2023equivariant}.

Following \cite{cheng2023equivariant} we use \textit{normalized mean absolute error} (NMAE) as our evaluation metric:
\begin{equation}
\text{NMAE} = \frac{\int_{\mathbb{R}^3} |\hat{\rho}(\mathbf{x}) - \rho(\mathbf{x})| d\mathbf{x}}{\int_{\mathbb{R}^3} |\rho(\mathbf{x})| d\mathbf{x}}
\end{equation} where $\hat{\rho}(\mathbf{x}), \rho(\mathbf{x})$ are the predicted (using Alg. \ref{alg1}) and ground truth electron densities, respectively. This is also used as the training objective. 

As baselines benchmarks we compare the following models performence as reported by \cite{cheng2023equivariant} and \cite{kim2024gaussian}. The baselines include: CNN \cite{ronneberger2015u}, interpolation networks (DeepDFT \cite{jorgensen2020deepdft}, DeepDFT2 \cite{J_rgensen_2022}, InfGCN\cite{cheng2023equivariant}, EGNN \cite{satorras2021n}, DimeNet \cite{gasteiger2020fast}, DimeNet++ \cite{gasteiger2020directional}), and neural operators (GNO \cite{li2020neural}, FNO \cite{li2020fourier}, LNO \cite{lu2021learning}) and the latest combination GPWNO\cite{kim2024gaussian}. For further details please refer to Appendix \ref{app:ds}.

\subsection{Results}
We present our main results in Table \ref{tab:gpwno_evaluation} demonstrating a significant improvement over multiple techniques, including recent ones such as \cite{kim2024gaussian}. Notably, the largest improvement margins are observed on the MD dataset, which contains highly variable geometries for the same molecule. These results highlight the importance of a correct geometric inductive bias in the model, which can significantly affect generalization. While our model performs well on the QM9 and MD datasets with relatively few parameters, achieving comparable results for the Cubic dataset required scaling the inner dimension and introducing an additional tensor product in Algorithm \ref{alg3}, resulting in a model with 120 million parameters. This was somewhat expected, as our atom-centered expansion is more localized and thus struggles with periodic systems. In contrast, on the MD dataset, we perform significantly better because our use of equivariant features—not just equivariant networks—provides a better inductive bias for modeling, grounded in a physics foundation. More details on the model parameters used for each dataset are provided in Appendix \ref{app:ds}.

\begin{table}[H]
\centering
\caption{We report the performance for QM9,MD, Cubic, dataset in NMAE (\%). The best number is highlighted in bold. The baseline results are following \cite{cheng2023equivariant},\cite{kim2024gaussian}}.
\label{tab:gpwno_evaluation}
\scriptsize
\begin{tabularx}{\textwidth}{l*{12}{>{\centering\arraybackslash}X}}
\toprule
Dataset & LNO & FNO & GNO & DmNet++ & DmNet & EGNN & DDFT2 & DDFT & CNN & InfGCN & GPWNO & \textbf{E3STO (Ours)} \\
\midrule
QM9 & 26.14 & 28.83 & 40.86 & 11.69 & 11.97 & 11.92 & 1.03 & 2.95 & 2.01 & 0.93 & 0.73 & \textbf{0.46} \\
\midrule
Ethanol & 43.17 & 31.98 & 82.35 & 14.24 & 13.99 & 13.90 & 8.83 & 7.34 & 13.97 & 8.43 & 4.00 & \textbf{1.47} \\
Benzene & 38.82 & 20.05 & 82.46 & 14.34 & 14.48 & 13.49 & 5.49 & 6.61 & 11.98 & 5.11 & 2.45 & \textbf{0.78} \\
Phenol & 60.70 & 42.98 & 66.69 & 12.99 & 12.93 & 13.59 & 7.00 & 9.09 & 11.52 & 5.51 & 2.68 & \textbf{0.70} \\
Resorcinol & 35.07 & 26.06 & 58.75 & 12.01 & 12.04 & 12.61 & 6.95 & 8.18 & 11.07 & 5.95 & 2.73 & \textbf{0.76} \\
Ethane & 77.14 & 26.31 & 71.12 & 12.95 & 13.11 & 15.17 & 6.36 & 8.31 & 14.72 & 7.01 & 3.67 & \textbf{1.45} \\
MDA & 47.22 & 34.58 & 84.52 & 16.79 & 18.71 & 12.37 & 10.68 & 9.31 & 18.52 & 10.34 & 5.32 & \textbf{1.49} \\
\midrule
Cubic & 46.33 & 48.08 & 53.55 & 12.18 & 12.51 & 11.74 & 10.37 & 14.08 & OOM & 5.18 & \textbf{4.89} & 5.1 \\

\bottomrule
\end{tabularx}
\label{tab:res}
\end{table}

\begin{figure}[H]
    \centering
    \includegraphics[width=\linewidth]{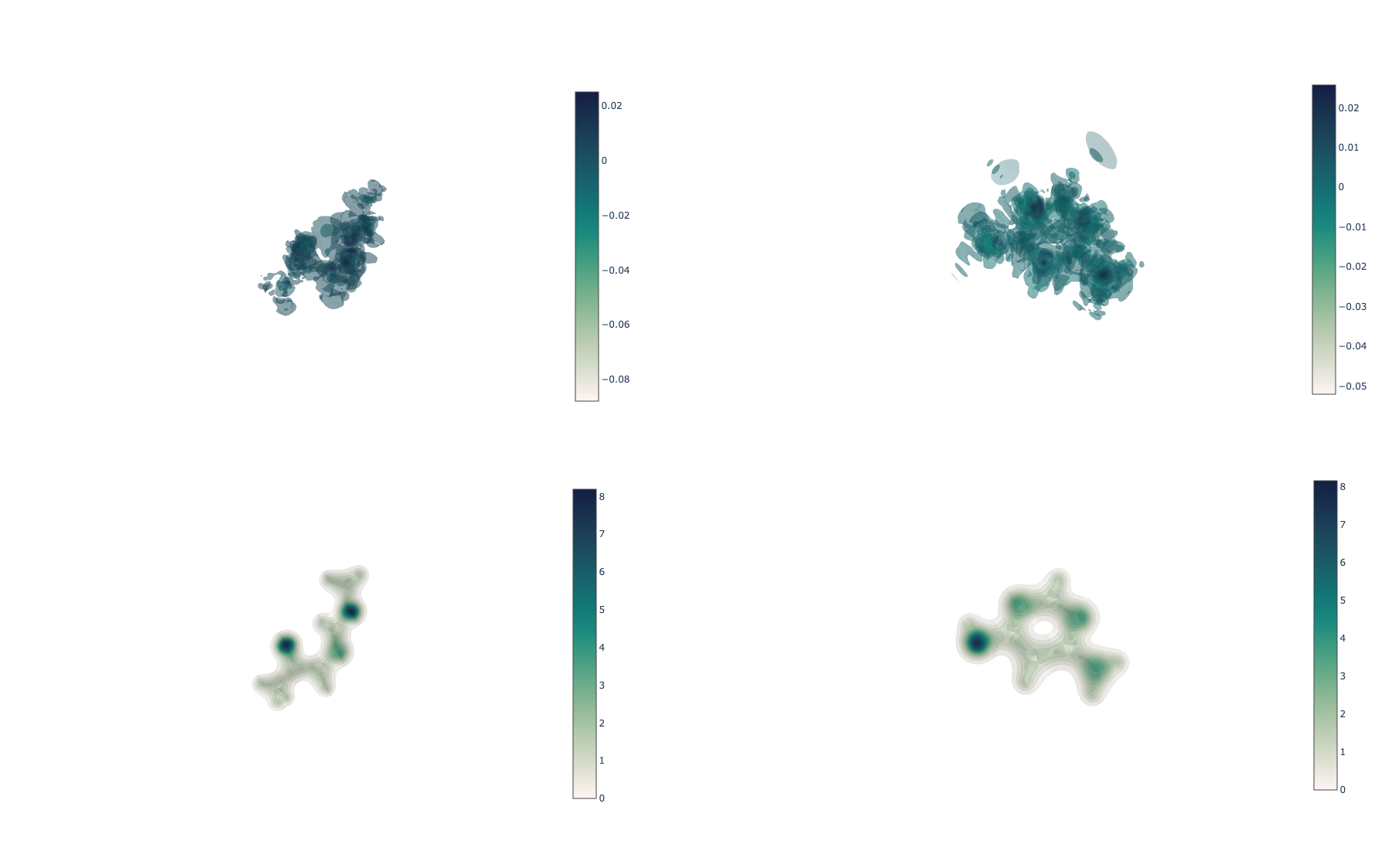}
    \caption{Top and bottom rows are the error and prediction of electronic density of two different molecules. Left and right errors are NMAE=0.47\% and 0.49\%.}
    \label{fig:res}
\end{figure}

Fig. \ref{fig:res} illustrates the predicted electronic density and the 3D error distribution. It shows that the relative error is small and primarily localized around the molecule. This indicates that our functional form, chosen for the molecular orbitals in Section 3.1, accurately describes the localized 3D electron distribution, as there appears to be no significant error further from the molecule. Additional visualization are available in Appendix \ref{app:plots}.

\subsection{Ablations}
First, we ablate the order of spherical harmonics used to represent and calculate the different parameters depending on the molecular geometry. In Table \ref{ab1}, we find an improvement up to $\ell=5$. This demonstrates the significance of the higher-order tensor representation while maintaining a similar number of parameters. This reflects the fact that molecules possess high degrees of symmetry, making it beneficial to model these symmetries directly. Furethermore, we ablate the functional form of our predicted electron density. To do that we set $\mathcal{C}_{n\ell}$, $f_{n\ell}$, and \textbf{MLP} terms to 1, while $\eta_{n\ell}$ and $\zeta_{n\ell}$, which act as exponents, are set to 0. Interestingly, in Table \ref{ab2}, we find an expected strong reliance on the orbital coefficients $\mathcal{C}_{n\ell}$ and the radial invariant function $f_{n\ell}$. However, we also see that $\eta_{n\ell}$ and $\zeta_{n\ell}$ do not have a strong effect on their own. We presume that due to the high expressiveness of $f_{n\ell}$, it is able to compensate for the lack of the exponential terms. These results further suggest that enabling a flexible and dynamic basis expansions which is dependent on the molecule and does not assume a predetermined form is beneficial for accuracy. Finally, we find that the inclusion of the \textbf{MLP} significantly improves accuracy. This improvement may be due to systemic errors in the dataset or the presence of a consistent background density that depends only on the atoms themselves and not their interactions.

\begin{table}[H]
\centering
\caption{NMAE (\%) and the parameter count of different model settings on the QM9 dataset.}
\begin{tabular}{ccccccccccccc}
\toprule
\textbf{Model} & $\ell_{max}=1$ & $\ell_{max}=2$ & $\ell_{max}=3$ & $\ell_{max}=4$ & \textbf{$\ell_{max}=5$} & $\ell_{max}=6$ \\
\midrule
\text{QM9 (\%)} & 6.67 & 2.58 & 1.65 & 0.70 & \textbf{0.46} & 0.47 \\
\midrule
\text{Parameters (M)} & 4.47 & 4.72 & 5.09 & 5.11 & 17.9 & 23.1 \\
\bottomrule
\end{tabular}
\label{ab1}

\end{table}

\begin{table}[H]
\centering
\caption{Ablation study on QM9. Performance in NMAE (\%) for the models missing one learnable feature of the basis expansion, i.e. All ablated models had the same number of parameters $5M$ and used $\ell_{max}=3$. The best number is highlighted in bold.}

\begin{tabular}{ccccccc}
\toprule
$\mathcal{C}_{n\ell}$ & $\eta_{n\ell}$ & $\zeta_{n\ell}$ & $f_{n\ell}$ & $\mathbf{MLP}$ & {\textbf{QM9}} \\
 & & & & & \\
\midrule
\textcolor{red}{\texttimes} & \checkmark & \checkmark & \checkmark & \checkmark & 10.04 \\
\checkmark & \textcolor{red}{\texttimes} & \checkmark & \checkmark & \checkmark & 1.70 \\
\checkmark & \checkmark & \textcolor{red}{\texttimes} & \checkmark & \checkmark & 1.68 \\
\checkmark & \checkmark & \checkmark & \textcolor{red}{\texttimes} & \checkmark & 24.54 \\
\checkmark & \checkmark & \checkmark & \checkmark & \textcolor{red}{\texttimes} & 1.81 \\
\checkmark & \checkmark & \checkmark & \checkmark & \checkmark & \textbf{1.65} \\
\bottomrule
\end{tabular}
\label{ab2}

\end{table}


\section{Limitations and Conclusion}
In this work, we developed an SE(3)-equivariant model based on Slater-type orbitals for predicting electronic densities, decoupling electron density prediction from the dependence on predefined basis functions. Our model demonstrated superior performance on small to medium-sized molecules, achieving significant improvements over state-of-the-art models on the MD dataset. Additionally, our results indicated the significance of flexibility in the radial functional form, opening avenues for future exploration. However, the model exhibited limitations in handling the cubic lattice structure, suggesting a need to reconsider the suitability of the atom-centered expansion approach for lattice-based systems. Furthermore, demonstrating generalizability to very large systems remains an open challenge, necessitating further investigation to ensure effective scaling. Despite these limitations, our work shows that leveraging the power of equivariance enables high-accuracy modeling of physical phenomena with relative ease compared to other approaches. The significant margin of improvement over the next best baseline on the MD dataset underscores the importance of robust equivariant representation. We hope this work advances the field closer to accurately approximating the physical and chemical properties of the natural world.

\begin{ack}
We would like to acknowledge Allan dos Santos Costa for helpful discussions.
\end{ack}

\bibliography{neurips_2024}
\bibliographystyle{unsrtnat}


\appendix

\section{Experimental Details}
\label{app:ds}

In this section, we provide more details about our dataset and train configuration. All the datasets contain the molecule graph of each species and its voxelized 3D electron charge density data on its pre-defined grid. We use the standard adam optimizer with learning rate of 0.001, without decay. We train each of the models for 100k steps, with batch size of 20. We use a single Nvidia A6000. Training takes 24hr-30hr for each. Our neural networks and code is written using Jax. Each experiment was repeated 3 times. 

\subsection*{QM9}
The dataset is provided by \cite{jorgensen2020deepdft}. The electron density of the dataset is calculated by VASP. To be specific, VASP with Perdew-Burke-Ernzerhof exchange correlation (PBE XC) functional and projector-augmented wave (PAW) is used. For this dataset we have used tensors of dimension 4 and $l$ up to 7. With 4 convolutional layers and up to 21M parameters.

\subsection*{MD}
The dataset is provided by \cite{bogojeski2020quantum}\cite{bogojeski2018efficient} and it is calculated by PBE XC functional with Quantum ESPRESSO using the PBE functional. For this dataset we have used tensors of dimension 4 and $l$ up to 6. With 4 convolutional layers, up to 5M parameters.

\subsection*{Cubic}
The densities are calculated with VASP using the projector augmented wave (PAW) method \cite{wang2022large}. As the crystal structure satisfies the periodic boundary condition (pbc), the volumetric data are given for a primitive cell with translation periodicity. We only focus on the total charge density and ignore the spin density. For this dataset we have used a model of dimension 8 and $l$ up to 2. With 4 convolutional layers and an extra tensor product resuting in 120M parameters.

\begin{table}[H]
    \centering
    \caption{Statistics of the dataset. We round up the number of grids and nodes to the nearest integer for readability.}
    \begin{tabular}{l*{3}{c}}
        \toprule
        Dataset & QM9 & Cubic & MD \\
        \midrule
        train/val/test split & 123835/50/1600 & 14421/1000/1000 & 1000(2000)/500(400) \\
        min/mean/max \#grid & 40/88/160 & 32/94/448 & 20/20/20 \\
        min/mean/max \#node & 3/18/29 & 1/10/64 & 8/11/14 \\
        min/mean/max length (Bohr) & 4.00/8.65/15.83 & 1.78/5.82/26.20 & 20/20/20 \\
        \# node type & 5 & 84 & 3 \\
        \bottomrule
    \end{tabular}
\end{table}

\section{Electron Density Prediction Visualization}
\label{app:plots}
\begin{figure}[H]
    \centering
    \resizebox{\textwidth}{!}{%
        \includegraphics{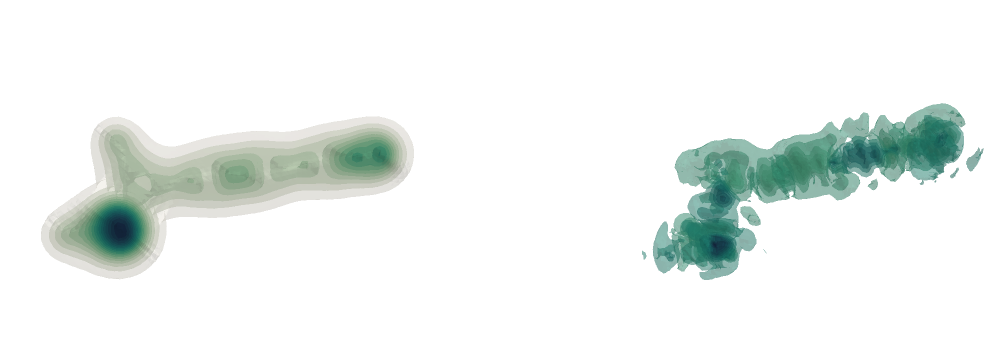}
    }
    \caption{Left: prediction, Right: error. NMAE=0.49\%}
    \label{fig:1}
\end{figure}
\begin{figure}[H]
    \centering
    \resizebox{\textwidth}{!}{%
        \includegraphics{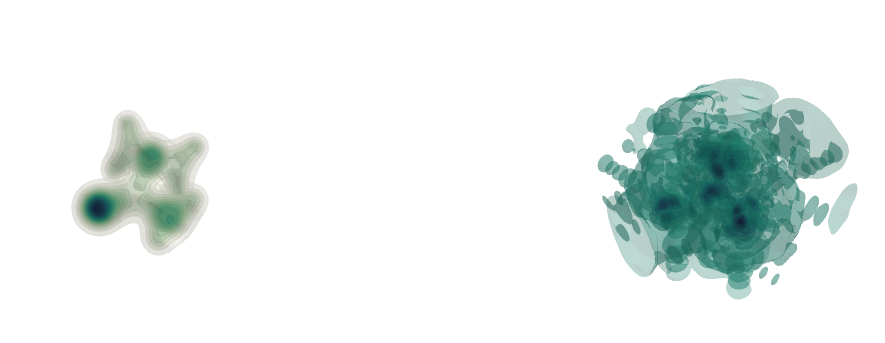}
    }
    \caption{Left: prediction, Right: error. NMAE=0.57\%}
    \label{fig:2}
\end{figure}
\begin{figure}[H]
    \centering
    \resizebox{\textwidth}{!}{%
        \includegraphics{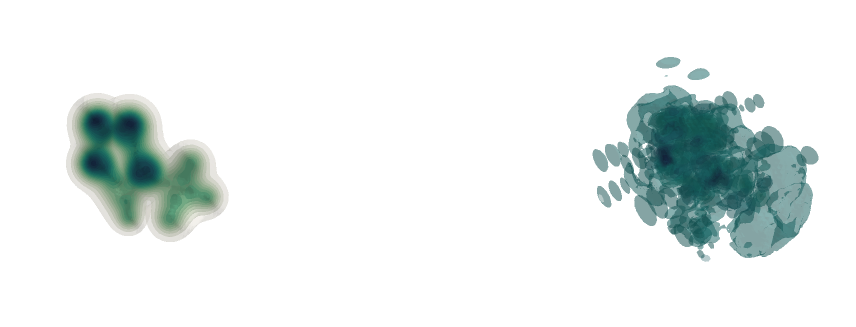}
    }
    \caption{Left: prediction, Right: error. NMAE=0.56\%}
    \label{fig:3}
\end{figure}
\begin{figure}[H]
    \centering
    \resizebox{\textwidth}{!}{%
        \includegraphics{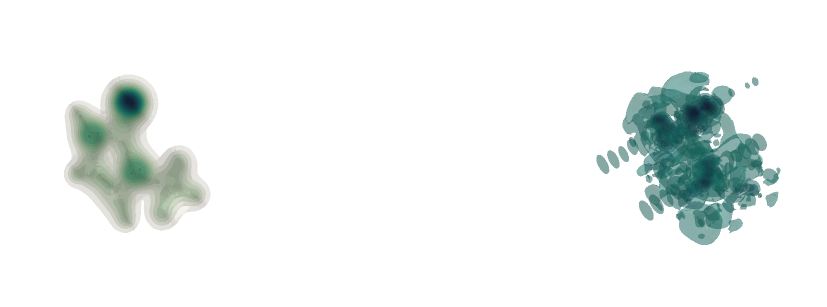}
    }
    \caption{Left: prediction, Right: error. NMAE=0.44\%}
    \label{fig:4}
\end{figure}
\begin{figure}[H]
    \centering
    \resizebox{\textwidth}{!}{%
        \includegraphics{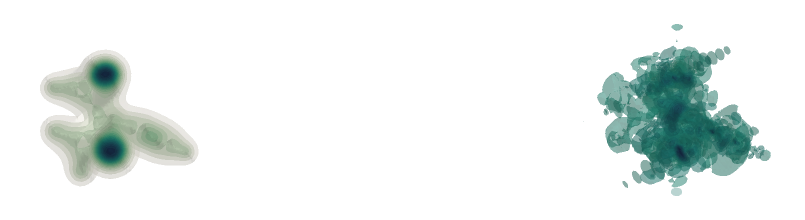}
    }
    \caption{Left: prediction, Right: error. NMAE=0.45\%}
    \label{fig:5}
\end{figure}
\begin{figure}[H]
    \centering
    \resizebox{\textwidth}{!}{%
        \includegraphics{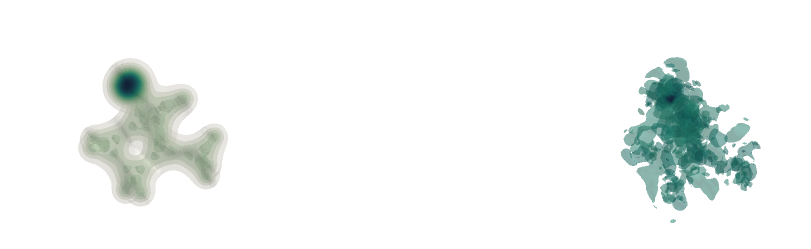}
    }
    \caption{Left: prediction, Right: error. NMAE=0.29\%}
    \label{fig:6}
\end{figure}
\begin{figure}[H]
    \centering
    \resizebox{\textwidth}{!}{%
        \includegraphics{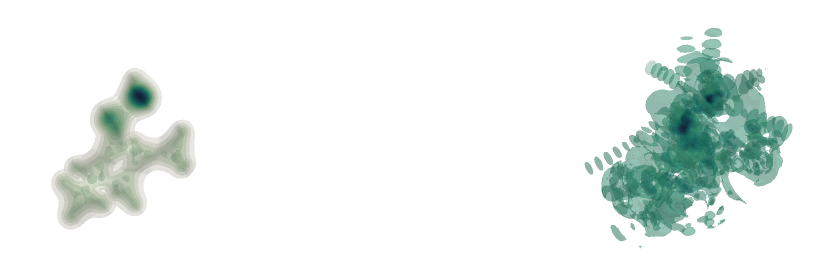}
    }
    \caption{Left: prediction, Right: error. NMAE=0.40\%}
    \label{fig:7}
\end{figure}

\end{document}